\documentclass[%
reprint,
nofootinbib,
nobibnotes,
amsmath,amssymb,
prl,
aps
]{revtex4-2}

\usepackage{graphicx}
\usepackage{dcolumn}
\usepackage{bm}

\usepackage{setspace}
\usepackage[utf8]{inputenc}

\usepackage[english]{babel}
\usepackage{hyphenat}
\newcommand{\rcite}[2][]{%
	ref.~%
	\ifx\relax#1\relax
	\cite{#2}%
	\else
	\cite[#1]{#2}%
	\fi
}
\newcommand{\rcites}[1]{refs.~\cite{#1}}

\usepackage{amsmath, amssymb, amsfonts, amsthm}
\usepackage{mathtools}
\usepackage[normalem]{ulem}

\theoremstyle{plain}

\newtheorem*{theorem*}{Theorem}

\usepackage{mathfixs}
\ProvideMathFix{autobold}
\ProvideMathFix{multskip}
\ProvideMathFix{frac,root,rootclose}

\allowdisplaybreaks[1]

\makeatletter
\def\mr@ignsp#1 {\ifx\:#1\@empty\else #1\expandafter\mr@ignsp\fi}%
\newcommand{\multiref}[1]{\begingroup
	\xdef\mr@no@sparg{\expandafter\mr@ignsp#1 \: }%
	\def\mr@comma{}%
	\@for\mr@refs:=\mr@no@sparg\do{\mr@comma\def\mr@comma{,}\ref{\mr@refs}}%
	\endgroup}
\renewcommand{\eqref}[1]{(\multiref{#1})}
\makeatother

\makeatletter
\newcommand{\namedref}[2]{\hyperref[#2]{#1~\ref*{#2}}}%
\newcommand{\namedreff}[2]{\hyperref[#2]{#1\,\ref*{#2}}}%

\makeatother

\newcommand{\eqn}[1]{eq.~\eqref{#1}}

\providecommand{\href}[2]{#2}

\usepackage[justification=justified,singlelinecheck=false,font=small,labelfont=bf]{caption}

\usepackage{graphbox}

\usepackage[compile=false,fonts]{mpostinl}
\begin{mposttex}[]
	\usepackage{amsmath,amssymb,amsfonts}
\end{mposttex}
\begin{mpostdef}
	input metaobj;
	
	def drawzigzag(expr pzz)=
	pair zz[];
	zz[1]:= point 0 of pzz ; zz[2]:= point 1 of pzz;
	nczigzag(zz[1])(zz[2]) "coilwidth(0.15xu)", "linewidth(1pt)", "linearc(.01xu)", "arrows(-)", "coilarmA(0)", "coilarmB(0)";
	enddef;
\end{mpostdef}

\begin{mpostdef}
	pair vpos[];
	pair epos[];
	pair ext[];
	pair exta[];
	path paths[];
	picture pic;
	picture pic[];
	picture savepic;
	path cpath;
	path hpath;
	xu:=1cm;
	yu:=1cm;
\end{mpostdef}

\begin{mpostdef}
	def pensize(expr s)=withpen pencircle scaled s enddef;
	def fillshape(expr p,ci,tb,cb)=
	fill p withcolor ci;
	draw p pensize(tb) withcolor cb;
	enddef;
	def filldot(expr z,s,ci)=
	fillshape(fullcircle scaled s shifted z, ci, 0.5pt, 0.0white);
	enddef;
	def filltrig(expr z,s,r,ci)=
	fillshape(((dir 60)--(dir 180)--(dir 300)--cycle) scaled s rotated r shifted z, ci, 0.5pt, 0.0white);
	enddef;
	def fillsqr(expr z,s,r,ci)=
	fillshape(((dir 45)--(dir 135)--(dir 225)--(dir 315)--cycle) scaled s rotated r shifted z, ci, 0.5pt, 0.0white);
	enddef;
	def drawcross(expr z,s,r,t,c)=
	draw ((-0.5,-0.5)--(+0.5,+0.5)) scaled s rotated r shifted z pensize(t) withcolor c;
	draw ((+0.5,-0.5)--(-0.5,+0.5)) scaled s rotated r shifted z pensize(t) withcolor c;
	enddef;
	
	def midarrow (expr p, t) =
	fill (arrowhead subpath(0,arctime(arclength(subpath (0,t) of p)+0.5ahlength) of p) of p) withcolor ecolor;
	enddef;
	
	def midarrowred (expr p, t) =
	fill (arrowhead subpath(0,arctime(arclength(subpath (0,t) of p)+0.5ahlength) of p) of p) withcolor ecolor;
	enddef;
	
	def midarrowreddark (expr p, t) =
	fill (arrowhead subpath(0,arctime(arclength(subpath (0,t) of p)+0.5ahlength) of p) of p) withcolor ecolordark;
	enddef;
	
	def midarrowblue (expr p, t) =
	fill (arrowhead subpath(0,arctime(arclength(subpath (0,t) of p)+0.5ahlength) of p) of p) withcolor bcolor;
	enddef;
	
	def midarrowbluedark (expr p, t) =
	fill (arrowhead subpath(0,arctime(arclength(subpath (0,t) of p)+0.5ahlength) of p) of p) withcolor bcolordark;
	enddef;	
	
	def midarrowgreen (expr p, t) =
	fill (arrowhead subpath(0,arctime(arclength(subpath (0,t) of p)+0.5ahlength) of p) of p) withcolor gcolor;
	enddef;
	
	def midarrowgreenInv (expr p, t) =
	fill (arrowhead subpath(arctime(arclength(subpath (0,t) of p)+0.5ahlength) of p,0) of p) withcolor gcolor;
	enddef;
	
	def dprop expr p=draw p pensize(3pt) withcolor feyncolor enddef;
	def dpropf expr p=draw p pensize(3pt) withcolor feyncolorfaint enddef;
	
	def drawprop expr p=draw p pensize(1.5pt) enddef;
	def drawline expr p=draw p pensize(0.5pt) enddef;
	def drawpos expr p=drawcross(p, 5pt, 0, 1.0pt, 0.0white) enddef;
	def drawvertex expr p=filldot(p, 5pt, 0.5white+0.5red) enddef;
	def drawblackdotscale expr p=filldot(p, 0.4*xu, black) enddef;
	def drawreddotscale expr p=filldot(p, 0.4*xu, red) enddef;
	def drawwhitedotscale expr p=filldot(p, 0.4*xu, white) enddef;
	def drawblackdot expr p=filldot(p, 8pt, black) enddef;
	def drawwhitedot expr p=filldot(p, 8pt, white) enddef;
	def drawlinearrow expr p=draw p pensize(0.5pt); midarrow (p,0.5); enddef;
	def drawproparrow expr p=draw p pensize(1.5pt); midarrow (p,0.5); enddef;
	def drawproparrowdouble expr p=draw p pensize(1.5pt); midarrow (p,0.65); midarrow (reverse p,0.65); enddef;
	def drawwhitedotmr (expr p, sz)=filldot(p, sz*xu, white) enddef;
	def drawblackdotmr (expr p, sz)=filldot(p, sz*xu, black) enddef;
	
	def framed(expr p)=fill bbox p withcolor white; draw bbox p pensize(0.8pt); draw p; enddef;
	def framedd(expr p)=fill bbox p withcolor 0.94white; draw bbox p pensize(1pt); draw p; enddef;
\end{mpostdef}
\begin{mpostdef}
	color ecolor; ecolor:=(.72,.03,.06);
	color bcolor; bcolor:=(.07,.05,.66);
	color gcolor; gcolor:=(.0,.44,.0);
	color ecolordark; ecolordark:=.3[ecolor,black];
	color bcolordark; bcolordark:=.3[bcolor,black];
	color feyncolor; feyncolor:=(.06,.4,.01);
	color feyncolorfaint; feyncolorfaint:=(.12,.8,.02);
	color arrowcolor; arrowcolor:=(0,0,0);
	color faint; faint:=(.7,0.7,0.7);
	color fgray; fgray:=(.7,0.7,0.7);
	vsize := 0.16xu;
	rad:= 2xu;
	cpath:= for i=0 upto 35: 
	dir (i*10) scaled rad ..
	endfor cycle;
	def vert(expr pos)=
	fill fullcircle scaled vsize shifted point pos of cpath withcolor white;
	draw halfcircle scaled vsize rotated (90+10*pos) shifted point pos of cpath withcolor ecolor pensize(1.3pt);
	enddef;
	def vertthick(expr pos)=
	fill fullcircle scaled vsize shifted point pos of cpath withcolor white;
	draw halfcircle scaled vsize rotated (90+10*pos) shifted point pos of cpath withcolor ecolor pensize(2.0pt);
	enddef;
	def vertcol(expr pos,col)=
	fill fullcircle scaled vsize shifted point pos of cpath withcolor white;
	draw halfcircle scaled vsize rotated (90+10*pos) shifted point pos of cpath withcolor col pensize(1.3pt);
	enddef;
	def drawbdry(expr startpos,endpos)=
	draw subpath(startpos,endpos) of cpath withcolor ecolor pensize(1.3pt);
	enddef;
	def drawbdryblack(expr startpos,endpos)=
	draw subpath(startpos,endpos) of cpath withcolor black pensize(1.3pt);
	enddef;
	def drawbdrythick(expr startpos,endpos)=
	draw subpath(startpos,endpos) of cpath withcolor ecolor pensize(2.0pt);
	enddef;
	def drawbdrythickblack(expr startpos,endpos)=
	draw subpath(startpos,endpos) of cpath withcolor black pensize(2.0pt);
	enddef;
	def drawbdrydashed(expr startpos,endpos)=
	draw subpath(startpos,endpos) of cpath withcolor ecolor dashed withdots scaled 0.5 pensize(1.3pt);
	enddef;
	def drawbdrydashedthick(expr startpos,endpos)=
	draw subpath(startpos,endpos) of cpath withcolor ecolor dashed withdots scaled 0.5 pensize(2.0pt);
	enddef;
	def vlabel(expr pos, vscal, tt)=
	vert(pos);
	label(tt,point pos of cpath scaled vscal);
	enddef;
	
	def vlabelcol(expr pos, vscal, tcol, vcol, tt)=
	vertcol(pos, vcol);
	label(tt,point pos of cpath scaled vscal) withcolor tcol;
	enddef;
	
	def ppath(expr p,spos,epos)=
	subpath(arctime(spos*arclength(p))of p,arctime(epos*arclength(p)) of p) of p
	enddef;
\end{mpostdef}

\begin{mpostdef}
	picture nicearrow; currentpicture:= nullpicture;
	ahlengthsave:=ahlength; ahlength:=12pt;
	drawarrow (0xu,0xu){dir 20}..(2xu,0xu) pensize(3pt) withcolor arrowcolor;
	ahlength:=ahlengthsave;
	nicearrow:= currentpicture; currentpicture:= nullpicture;
\end{mpostdef}

\begin{mpostdef}
	eNum:= 2.718281828459045235;
	numeric startangle, lsdist, scalefactor, stepangle;
	path ls;
	
	vardef logspiralder(expr lspos, endpoint, maxangle, stepnumber, enddir) =
	startangle := ((angle(endpoint-lspos) - (maxangle mod 360)) mod 360); 
	stepangle := (1.0maxangle)/(1.0stepnumber);
	lsdist := length(endpoint-lspos); 
	scalefactor := mlog(lsdist+1.0)/(1.0*maxangle)/256;
	ls:=lspos for t = startangle step stepangle until (maxangle + startangle + 0.02): .. (((mexp(256((t-startangle)*scalefactor))-1.0) * cosd(t),(mexp(256((t-startangle)*scalefactor))-1.0)  * sind(t))+lspos) endfor;
	ls:= subpath(1,(length ls) - 1) of ls;
	ls:= ls .. {dir enddir}endpoint;
	ls
	enddef;
\end{mpostdef}

\begin{mpostdef}
def roundedrectangle(expr xa, ya, xb, yb, r) =
  (xa + r, ya) 
  -- (xb - r, ya) {dir 0} .. {dir 90} (xb, ya + r) 
  -- (xb, yb - r) {dir 90} .. {dir 180} (xb - r, yb) 
  -- (xa + r, yb) {dir 180} .. {dir 270} (xa, yb - r) 
  -- (xa, ya + r) {dir 270} .. {dir 0} cycle
enddef;

def SchottkyBackground(expr xoffset, yoffset, w, h) =
fregion := 0.2xu;
fsteps := 80;

for ii = 0 upto fsteps:
  numeric margin, g;
  margin := ii / fsteps * fregion;
  gcol := 1.0 - 0.3 * (ii / fsteps); 

  draw roundedrectangle(margin+xoffset, margin+yoffset,w - margin+xoffset, h-margin+yoffset, fregion * (1-ii / fsteps) ) withcolor (gcol, gcol, gcol) pensize( fregion / (fsteps - 1));
endfor
  fill roundedrectangle(margin+xoffset, margin+yoffset,w - margin+xoffset, h-margin+yoffset, 0 ) withcolor (0.7,0.7,0.7);
enddef;

def SchottkyAxes(expr xstartpos, xendpos, ystartpos, yendpos) = 
  drawarrow xstartpos-- xendpos pensize (1.0pt);  
  drawarrow ystartpos -- yendpos pensize (1.0pt);  
enddef;

def SchottkyCircle(expr centerPos, rad, col, bcol) = 
  fill fullcircle scaled (2*rad) shifted centerPos withcolor bcol; 
  draw fullcircle scaled (2*rad) shifted centerPos withcolor col pensize(0.7pt);
enddef;

def SchottkyCircleNoFilling(expr centerPos, rad, col) = 
  draw fullcircle scaled (2*rad) shifted centerPos withcolor col pensize(0.7pt);
enddef;

def SchottkyACircle(expr centerPos, rad, col, bcol) = 
  fill fullcircle scaled (2*rad) shifted centerPos withcolor bcol;
  drawarrow reverse(fullcircle) scaled (2*rad) shifted centerPos withcolor col pensize(0.7pt);
enddef;

def SchottkyCircleDashed(expr centerPos, rad, col, bcol) = 
	fill fullcircle scaled (2*rad) shifted centerPos withcolor bcol;
	draw fullcircle scaled (2*rad) shifted centerPos withcolor col pensize(0.7pt) dashed evenly;
enddef;

def SchottkyCircleDashedNoFilling(expr centerPos, rad, col) = 
	draw fullcircle scaled (2*rad) shifted centerPos withcolor col pensize(0.7pt) dashed evenly;
enddef;

def SchottkyCross(expr PosA, Sangle, col) = 
  transform ST;
  ST := identity rotated (Sangle + 45 + 15) shifted PosA; 
  draw ((-0.05xu,0xu)--(0.05xu,0xu)) transformed ST withcolor col;
  draw ((0,-0.05xu)--(0,0.05xu)) transformed ST withcolor col;
enddef;

def SchottkyCrossS(expr PosA, Sangle, size, col) = 
  transform ST;
  ST := identity rotated (Sangle + 45 + 15) shifted PosA; 
  draw ((-size,0xu)--(size,0xu)) transformed ST withcolor col;
  draw ((0,-size)--(0,size)) transformed ST withcolor col;
enddef;

def SchottkySquare(expr PosA, Sangle) = 
	transform ST;
	ST := identity rotated (Sangle + 45 + 15) shifted PosA; 
	draw ((-0.03xu,-0.03xu)--(0.03xu,-0.03xu)--(0.03xu,0.03xu)--(-0.03xu,0.03xu)--(-0.03xu,-0.03xu)) transformed ST;
enddef;

def SchottkyPathArrow(expr PosA,PosB) =
  path SPA;
  numeric SPAdir;
  SPAdir := angle(PosB-PosA);
  SPA := PosA {dir (SPAdir-15)} .. {dir (SPAdir+15)} PosB;
  drawarrow subpath (0.04 * length SPA, 0.96 * length SPA) of SPA pensize(0.7pt);
  SchottkyCross (PosB, SPAdir, (0,0,0));
enddef;

def SchottkyPathArrowAlt(expr PosA,PosB) =
  path SPA;
  numeric SPAdir;
  SPAdir := angle(PosB-PosA);
  SPA := PosA {dir (SPAdir+15)} .. {dir (SPAdir-15)} PosB;
  drawarrow subpath (0.04 * length SPA, 0.96 * length SPA) of SPA pensize(0.7pt);
  SchottkyCross (PosB, SPAdir, (0,0,0));
enddef;
\end{mpostdef}

\begin{mpostdef}
	path quartercircle;
	quartercircle := subpath (0, 2) of fullcircle;
\end{mpostdef}

\begin{mpostfig}[label=associatorrelation]
	Retau := 1.3xu;
	Imtau := 4.0xu;
	maxX := 6.0xu;
	maxY := 4.5xu;
	markedPointX := 2.7xu;
	markedPointY := 1.7xu;
	
	ahlength := 4bp;
	drawarrow (-0.1xu, 0xu) -- (maxX, 0xu) pensize (1pt);  
	drawarrow (0xu, -0.1xu) -- (0xu, maxY) pensize (1pt);  
	label.bot(btex $\textrm{Re}(z)$ etex, (maxX-0.5xu, 0xu));
	label.lft(btex $\textrm{Im}(z)$ etex, (0xu, maxY-0.5xu));
	
	ahlength := 0.2cm;
	draw (Retau,Imtau) -- (maxX-0.5xu,Imtau) pensize (1pt) withcolor ecolor;
	midarrowred((maxX-0.5xu,Imtau)--(Retau,Imtau),0.5);
	draw (0.0xu,0.0xu) -- (maxX-Retau-0.5xu,0.0xu) pensize (1pt) withcolor ecolor;
	midarrowred((0.0xu,0.0xu)--(maxX-Retau-0.5xu,0.0xu),0.5);
	draw (0.0xu,0.0xu) -- (Retau,Imtau) pensize (1pt) withcolor bcolor;
	midarrowblue((Retau,Imtau)--(0.0xu,0.0xu),0.5);
	draw (maxX-0.5xu,Imtau) -- (maxX-Retau-0.5xu,0.0xu) pensize (1pt) withcolor bcolor;
	midarrowblue((maxX-Retau-0.5xu,0.0xu)--(maxX-0.5xu,Imtau),0.5);
	
	label.top(btex $\tau$ etex, (Retau,Imtau));
	label.bot(btex $1$ etex, (maxX-Retau-0.5xu, 0.0xu));
	label.top(btex $1+\tau$ etex, (maxX-0.5xu, Imtau));
	
	label.bot(btex $\mathfrak{A}$ etex, (0.5 * (maxX-Retau-0.5xu),0.0xu));
	label.rt(btex $\mathfrak{B}$ etex, (maxX-0.5*Retau-0.5xu,0.5*Imtau));
	label.top(btex $\mathfrak{A}^{-1}$ etex, (0.5 * (maxX+Retau-0.5xu),Imtau));
	label.rt(btex $\mathfrak{B}^{-1}$ etex, (0.5*Retau,0.5*Imtau));
	
	dotlabeldiam:=4pt;
	dotlabel.top (btex $ $ etex, (markedPointX,markedPointY));

	ahlength := 0.2cm;
	cycleRadius= 1.0xu;
	draw fullcircle xscaled (cycleRadius) yscaled (cycleRadius) shifted (markedPointX,markedPointY) pensize(1pt) withcolor gcolor;	
	midarrowgreen(fullcircle xscaled (cycleRadius) yscaled (cycleRadius) shifted (markedPointX,markedPointY),0.5);
	label.rt(btex $\mathfrak{C}$ etex, (markedPointX+0.5xu,markedPointY+0.2xu));
	pic[1]:=currentpicture;
	currentpicture:=nullpicture;
	
	draw fullcircle xscaled 8xu yscaled 3.8xu pensize (1pt);
	draw (-2xu,0.06xu){dir -20}..(2xu,0.06xu) pensize (1.0pt);
	draw (-1.6xu,-0.04xu){dir 25}..(1.6xu,-0.04xu) pensize (1.0pt);
	path apath; apath = (0xu,-0.3xu) .. controls (-0.2xu,-0.36xu) and (-0.2xu,-1.84xu) .. ( 0xu,-1.9xu);
	path bpath; bpath = (-4xu,0xu) ..controls (-4xu,-1.5xu) and (4xu,-1.5xu) .. (4xu,0xu);
	draw (-4xu,0xu) ..controls (-4xu,1.5xu) and (4xu,1.5xu) .. ( 4xu,0xu) dashed evenly pensize (1.pt) withcolor bcolor;
	draw (0xu,-0.3xu) .. controls (0.2xu,-0.36xu) and (0.2xu,-1.84xu) .. ( 0xu,-1.9xu) dashed evenly pensize (1.pt) withcolor ecolor;
	draw fullcircle xscaled 8xu yscaled 3.8xu pensize (1pt);
	draw (-2xu,0.06xu){dir -20}..(2xu,0.06xu) pensize (1.0pt);
	draw bpath pensize (1.pt) withcolor bcolor;
	draw apath pensize (1.pt) withcolor ecolor;
	dotlabeldiam:=4pt;
	dotlabel.bot (btex $ $ etex, (1.7xu,-0.6xu));
	path cpath; cpath = fullcircle xscaled 0.7xu yscaled 0.6xu rotated (10) shifted (1.69xu,-0.6xu);
	draw cpath withcolor gcolor pensize (1pt);
	ahlength := 0.2cm;
	midarrowblue(bpath,0.4);
	midarrowred(apath,0.4);
	midarrowgreen(cpath,0.5);
	label.lft(btex $\mathfrak{A}$ etex, point 0.4 of apath);
	label.top(btex $\mathfrak{B}$ etex, point 0.4 of bpath);
	label.rt(btex $\mathfrak{C}$ etex, point 0.4 of cpath);
	pic[2]:=currentpicture;
	currentpicture:=nullpicture;
	
	draw pic[1] shifted (0xu,0xu);
	draw pic[2] shifted (10.8xu,2xu);
\end{mpostfig}

\begin{mpostfig}[label=ellipticassociators]
	Retau := 1.3xu;
	Imtau := 4.0xu;
	maxX := 6.0xu;
	maxY := 4.5xu;
	markedPointX := 2.7xu;
	markedPointY := 1.7xu;
	
	ahlength := 4bp;
	drawarrow (-0.1xu, 0xu) -- (maxX, 0xu) pensize (1pt);  
	drawarrow (0xu, -0.1xu) -- (0xu, maxY) pensize (1pt);  
	label.bot(btex $\textrm{Re}(z)$ etex, (maxX-0.5xu, 0xu));
	label.lft(btex $\textrm{Im}(z)$ etex, (0xu, maxY-0.5xu));
	
	ahlength := 0.2cm;
	phaseRadius := 0.5xu;
	associatorPenSize := 1.7pt;
	draw (Retau,Imtau) -- (maxX-0.5xu,Imtau) pensize (1pt) withcolor ecolor;
	draw (0.0xu,0.0xu) -- (maxX-Retau-0.5xu,0.0xu) pensize (1pt) withcolor ecolor;
	draw (0.5*phaseRadius,0.0xu) -- (maxX-Retau-0.5xu-0.5*phaseRadius,0.0xu) pensize (associatorPenSize) withcolor ecolor;
	midarrowred((0.0xu,0.0xu)--(maxX-Retau-0.5xu,0.0xu),0.5);
	draw (0.0xu,0.0xu) -- (Retau,Imtau) pensize (1pt) withcolor bcolor;
	eps := 0.5*phaseRadius/sqrt(Retau**2+Imtau**2);
	draw (maxX-0.5xu,Imtau) -- (maxX-Retau-0.5xu,0.0xu) pensize (1pt) withcolor bcolor;
	draw (eps*4.2xu,0.0xu) -- (Retau+eps*4.2xu,Imtau) pensize (associatorPenSize) withcolor bcolor;
	draw halfcircle scaled phaseRadius shifted (maxX-Retau-0.5xu,0.0xu) pensize(associatorPenSize) withcolor ecolor;
	midarrowblue((eps*4.2xu,0.0xu)--(Retau+eps*4.2xu,Imtau),0.52);
	
	label.lft(btex $\tau$ etex, (Retau,Imtau));
	label.bot(btex $1$ etex, (maxX-Retau-0.5xu, 0.0xu));
	label.top(btex $1+\tau$ etex, (maxX-0.5xu, Imtau));
	
	label.bot(btex $A(\tau)$ etex, (0.5 * (maxX-Retau-0.5xu),0.0xu));
	label.rt(btex $B(\tau)$ etex, (0.5*Retau+eps*4.2xu,0.5*Imtau));
	pic[1]:=currentpicture;
	currentpicture:=nullpicture;
	
	draw fullcircle xscaled 8xu yscaled 3.8xu pensize (1pt);
	draw (-2xu,0.06xu){dir -20}..(2xu,0.06xu) pensize (1.0pt);
	draw (-1.6xu,-0.04xu){dir 25}..(1.6xu,-0.04xu) pensize (1.0pt);
	draw (-4xu,0xu) ..controls (-4xu,1.5xu) and (4xu,1.5xu) .. ( 4xu,0xu) dashed evenly pensize (0.75pt) withcolor bcolor;
	draw (0xu,-0.3xu) .. controls (0.2xu,-0.36xu) and (0.2xu,-1.84xu) .. ( 0xu,-1.9xu) dashed evenly pensize (0.75pt) withcolor ecolor;
	draw (-4xu,0xu) ..controls (-4xu,-1.5xu) and (4xu,-1.5xu) .. (4xu,0xu) pensize (0.75pt) withcolor bcolor;
	draw (0xu,-0.3xu) .. controls (-0.2xu,-0.36xu) and (-0.2xu,-1.84xu) .. ( 0xu,-1.9xu) pensize (0.75pt) withcolor ecolor;
	xwidth := 4xu;
	draw (-xwidth,0.0xu) ..controls (-xwidth,1.5xu) and (xwidth,1.5xu) .. (xwidth,0.0xu) dashed evenly pensize (1.5pt) withcolor bcolor;
	draw (0.0xu,-0.3xu) .. controls (0.2xu,-0.36xu) and (0.2xu,-1.84xu) .. (0.0xu,-1.9xu) dashed evenly pensize (1.5pt) withcolor ecolor;
	draw (-2xu,0.06xu){dir -20}..(2xu,0.06xu) pensize (1.0pt);
	draw (-4xu,0xu) ..controls (-4xu,-1.5xu) and (4xu,-1.5xu) .. (4xu,0xu) pensize (0.75pt) withcolor bcolor;
	draw fullcircle xscaled 8xu yscaled 3.8xu pensize (1pt);
	path bpath; bpath = (-xwidth,0.0xu) ..controls (-xwidth,-1.5xu) and (xwidth,-1.5xu) .. (xwidth,0.0xu);
	draw subpath (0,.4555) of bpath pensize (1.5pt) withcolor bcolor;
	midarrowblue(bpath,0.3);
	draw subpath (0.535,1) of bpath pensize (1.5pt) withcolor bcolor;
	draw subpath (0,-4.0) of fullcircle scaled 0.77xu shifted (-0.15xu,-1.12xu) pensize (1.5pt) withcolor bcolor dashed withdots scaled 0.4;
	path rpath; rpath = (0.0xu,-0.3xu) .. controls (-0.2xu,-0.36xu) and (-0.2xu,-1.84xu) .. (0.0xu,-1.9xu);
	draw subpath (0,0.41) of rpath pensize (1.5pt) withcolor ecolor;
	draw subpath (0.73,1) of rpath pensize (1.5pt) withcolor ecolor;
	midarrowred(rpath,0.3);
	draw subpath (-1.95,1.95) of fullcircle scaled 0.43xu shifted (-0.15xu,-1.12xu) pensize (1.5pt) withcolor ecolor;
	dotlabeldiam:=4pt;
	dotlabel.bot (btex $ $ etex, point .51 of ((0xu,-0.3xu) .. controls (-0.2xu,-0.36xu) and (-0.2xu,-1.84xu) .. ( 0xu,-1.9xu)));
	pic[2]:=currentpicture;
	currentpicture:=nullpicture;
	
	draw pic[1] shifted (0xu,0xu);
	draw pic[2] shifted (10.8xu,2xu);
\end{mpostfig}

\begin{mpostfig}[label=ellipticsvcondition]
	Retau := 1.3xu;
	Imtau := 4.0xu;
	maxX := 6.0xu;
	maxY := 4.5xu;
	markedPointX := 2.7xu;
	markedPointY := 1.7xu;
	
	ahlength := 4bp;
	drawarrow (-0.1xu, 0xu) -- (maxX, 0xu) pensize (1pt);  
	drawarrow (0xu, -0.1xu) -- (0xu, maxY) pensize (1pt);  
	label.bot(btex $\textrm{Re}(z)$ etex, (maxX-0.5xu, 0xu));
	label.lft(btex $\textrm{Im}(z)$ etex, (0xu, maxY-0.5xu));
	
	ahlength := 0.2cm;
	phaseRadius := 0.5xu;
	associatorPenSize := 1.7pt;
	draw (Retau,Imtau) -- (maxX-0.5xu,Imtau) pensize (1pt) withcolor ecolor;
	draw (0.0xu,0.0xu) -- (maxX-Retau-0.5xu,0.0xu) pensize (1pt) withcolor ecolor;
	draw (0.0xu,0.0xu) -- (Retau,Imtau) pensize (1pt) withcolor bcolor;
	draw (maxX-0.5xu,Imtau) -- (maxX-Retau-0.5xu,0.0xu) pensize (1pt) withcolor bcolor;
	
	draw (0.5*phaseRadius,0.1xu) -- (maxX-Retau-0.5xu-0.5*phaseRadius,0.1xu) pensize (associatorPenSize) withcolor ecolor;
	draw subpath (8,11.45) of fullcircle scaled phaseRadius shifted (maxX-Retau-0.5xu,0.0xu) pensize(associatorPenSize) withcolor ecolor;
	draw subpath (4.55,8) of fullcircle scaled phaseRadius shifted (maxX-Retau-0.5xu,0.0xu) pensize(associatorPenSize) withcolor ecolordark;
	draw (maxX-Retau-0.5xu-0.5*phaseRadius,-0.1xu) -- (0.5*phaseRadius,-0.1xu) pensize (associatorPenSize) withcolor ecolordark;
	midarrowred((0.1xu,0.1xu)--(maxX-Retau-0.5xu,0.1xu),0.5);
	midarrowreddark((maxX-Retau-0.5xu,-0.1xu)--(0.1xu,-0.1xu),0.5);
	eps := 0.5*phaseRadius/sqrt(Retau**2+Imtau**2);
	draw subpath (eps,1) of ((0.15xu,0.0xu) -- (Retau+0.15xu,Imtau)) pensize (associatorPenSize) withcolor bcolor;
	draw subpath (1,eps) of ((-0.15xu,0.0xu) -- (Retau-0.15xu,Imtau)) pensize (associatorPenSize) withcolor bcolordark;
	draw subpath (2.,4.5) of fullcircle scaled 0.3xu shifted (-sqrt(Retau*Retau+Imtau*Imtau),0.0xu) rotated (180+72) pensize (associatorPenSize) withcolor bcolor dashed withdots scaled 0.4;
	draw subpath (4.5,6.5) of fullcircle scaled 0.3xu shifted (-sqrt(Retau*Retau+Imtau*Imtau),0.0xu) rotated (180+72) pensize (associatorPenSize) withcolor bcolordark dashed withdots scaled 0.4;
	midarrowblue((0.15xu,0.0xu)--(Retau+0.15xu,Imtau),0.5);
	midarrowbluedark((Retau-0.15xu,Imtau)--(-0.15xu,0.0xu),0.5);
	
	label.lft(btex $\tau$ etex, (Retau-0.2xu,Imtau));
	label.bot(btex $1$ etex, (maxX-Retau-0.5xu, -0.3xu));
	label.top(btex $1+\tau$ etex, (maxX-0.5xu, Imtau));
	
	draw (0xu, 1.8xu) -- (0xu, 2.2xu) pensize (5pt) withcolor white;
	
	label.top(btex $A_{\mathcal{Y}}$ etex, (0.5 * (maxX-Retau-0.5xu),0.1xu));
	label.bot(btex $\mathrm{R}(\overline{A_{\mathcal{Y}'}})$ etex, (0.5 * (maxX-Retau-0.5xu),-0.1xu));
	label.rt(btex $B_{\mathcal{Y}}$ etex, (0.5*Retau+0.15xu,0.5*Imtau));
	label.lft(btex $\mathrm{R}(\overline{B_{\mathcal{Y}'}})$ etex, (0.5*Retau-0.15xu,0.5*Imtau));
	pic[1]:=currentpicture;
	currentpicture:=nullpicture;
	
	draw fullcircle xscaled 8xu yscaled 3.8xu pensize (1pt);
	draw (-2xu,0.06xu){dir -20}..(2xu,0.06xu) pensize (1.0pt);
	draw (-1.6xu,-0.04xu){dir 25}..(1.6xu,-0.04xu) pensize (1.0pt);
	draw (-4xu,0xu) ..controls (-4xu,1.5xu) and (4xu,1.5xu) .. ( 4xu,0xu) dashed evenly pensize (0.75pt) withcolor bcolor;
	draw (0xu,-0.3xu) .. controls (0.2xu,-0.36xu) and (0.2xu,-1.84xu) .. ( 0xu,-1.9xu) dashed evenly pensize (0.75pt) withcolor ecolor;
	draw (-4xu,0xu) ..controls (-4xu,-1.5xu) and (4xu,-1.5xu) .. (4xu,0xu) pensize (0.75pt) withcolor bcolor;
	draw (0xu,-0.3xu) .. controls (-0.2xu,-0.36xu) and (-0.2xu,-1.84xu) .. ( 0xu,-1.9xu) pensize (0.75pt) withcolor ecolor;
	xwidth := 3.98xu;
	draw subpath (0.03,0.97) of ((-xwidth,0.1xu) ..controls (-xwidth,1.6xu) and (xwidth,1.6xu) .. (xwidth,0.1xu)) dashed evenly pensize (1.5pt) withcolor bcolordark;
	draw subpath (0.03,0.97) of ((-xwidth,-0.1xu) ..controls (-xwidth,1.4xu) and (xwidth,1.4xu) .. (xwidth,-0.1xu)) dashed evenly pensize (1.5pt) withcolor bcolor;
	draw (0.1xu,-0.3xu) .. controls (0.3xu,-0.36xu) and (0.3xu,-1.84xu) .. (0.1xu,-1.9xu) dashed evenly pensize (1.5pt) withcolor ecolor;
	draw (-0.1xu,-0.3xu) .. controls (0.1xu,-0.36xu) and (0.1xu,-1.84xu) .. (-0.1xu,-1.9xu) dashed evenly pensize (1.5pt) withcolor ecolordark;
	draw fullcircle xscaled 8xu yscaled 3.8xu pensize (1pt);
	draw (-2xu,0.06xu){dir -20}..(2xu,0.06xu) pensize (1.0pt);
	path bpathu; bpathu = (-xwidth,0.1xu) ..controls (-xwidth,-1.4xu) and (xwidth,-1.4xu) .. (xwidth,0.1xu);
	draw subpath (0,.459) of bpathu pensize (1.5pt) withcolor bcolordark;
	draw subpath (0.53,1) of bpathu pensize (1.5pt) withcolor bcolordark;
	path bpathl; bpathl = (-xwidth,-0.1xu) ..controls (-xwidth,-1.6xu) and (xwidth,-1.6xu) .. (xwidth,-0.1xu);
	draw subpath (0,.459) of bpathl pensize (1.5pt) withcolor bcolor;
	draw subpath (0.53,1) of bpathl pensize (1.5pt) withcolor bcolor;
	draw subpath (-3.55,0) of fullcircle scaled 0.7xu shifted (-0.15xu,-1.12xu) pensize (1.5pt) withcolor bcolor dashed withdots scaled 0.3;
	draw subpath (0,3.57) of fullcircle scaled 0.7xu shifted (-0.15xu,-1.12xu) pensize (1.5pt) withcolor bcolordark dashed withdots scaled 0.3;
	path rpathr; rpathr = (0.1xu,-0.3xu) .. controls (-0.1xu,-0.36xu) and (-0.1xu,-1.84xu) .. (0.1xu,-1.9xu);
	draw subpath (0,0.41) of rpathr pensize (1.5pt) withcolor ecolor;
	draw subpath (0.72,1) of rpathr pensize (1.5pt) withcolor ecolor;
	draw (-4xu,0xu) ..controls (-4xu,-1.5xu) and (4xu,-1.5xu) .. (4xu,0xu) pensize (0.75pt) withcolor bcolor;
	draw subpath (-5.5,-1.98) of fullcircle scaled 0.5xu shifted (-0.15xu,-1.12xu) pensize (1.5pt) withcolor ecolordark;
	draw subpath (-1.98,1.46) of fullcircle scaled 0.5xu shifted (-0.15xu,-1.12xu) pensize (1.5pt) withcolor ecolor;
	path rpathl; rpathl = (-0.1xu,-0.3xu) .. controls (-0.3xu,-0.36xu) and (-0.3xu,-1.84xu) .. (-0.1xu,-1.9xu);
	draw subpath (0,0.4) of rpathl pensize (1.5pt) withcolor ecolordark;
	draw subpath (0.72,1) of rpathl pensize (1.5pt) withcolor ecolordark;
	dotlabeldiam:=4pt;
	dotlabel.bot (btex $ $ etex, point .51 of ((0xu,-0.3xu) .. controls (-0.2xu,-0.36xu) and (-0.2xu,-1.84xu) .. ( 0xu,-1.9xu)));
	midarrowblue(bpathl,0.4);
	midarrowbluedark(reverse bpathu,0.6);
	midarrowred(rpathr,0.3);
	midarrowreddark(reverse rpathl,0.75);
	pic[2]:=currentpicture;
	currentpicture:=nullpicture;
	
	draw pic[1] shifted (0xu,0xu);
	draw pic[2] shifted (10.8xu,2xu);
\end{mpostfig}

\begin{mpostfig}[label=genustwo]
	
	draw (-3xu,-0.15xu) .. controls (-2.9xu,-0.18xu) and (-2.9xu,-0.92xu) .. ( -3xu,-0.95xu) dashed evenly pensize (1.0pt) withcolor ecolor;
	draw ((-4.2xu,0xu) ..controls (-4.2xu,0.5xu) and (-1.8xu,0.5xu) .. (-1.8xu,0xu)) shifted (0xu,0.05xu) pensize (1.0pt) withcolor bcolor;
	draw ((-4.2xu,0xu) ..controls (-4.2xu,-0.5xu) and (-1.8xu,-0.5xu) .. (-1.8xu,0xu)) shifted (0xu,0.05xu) pensize (1.0pt) withcolor bcolor;
	midarrowblue ((-4.2xu,0.05xu) ..controls (-4.2xu,-0.45xu) and (-1.8xu,-0.45xu) .. (-1.8xu,0.05xu),0.6);
	draw (-4xu,0.03xu){dir -20}..(-2xu,0.03xu) pensize (1.0pt);
	draw (-3.8xu,-0.02xu){dir 25}..(-2.2xu,-0.02xu) pensize (1.0pt);
	draw (-3xu,-0.15xu) .. controls (-3.1xu,-0.18xu) and (-3.1xu,-0.92xu) .. ( -3xu,-0.95xu) pensize (1.0pt) withcolor ecolor;
	midarrowred ((-3xu,-0.15xu) .. controls (-3.1xu,-0.18xu) and (-3.1xu,-0.92xu) .. ( -3xu,-0.95xu),0.65);
	
	draw (0xu,-0.15xu) .. controls (0.1xu,-0.18xu) and (0.1xu,-0.92xu) .. ( 0xu,-0.95xu) dashed evenly pensize (1.0pt) withcolor ecolor;
	draw ((-1.2xu,0xu) ..controls (-1.2xu,0.5xu) and (1.2xu,0.5xu) .. (1.2xu,0xu)) shifted (0xu,0.05xu) pensize (1.0pt) withcolor bcolor;
	draw ((-1.2xu,0xu) ..controls (-1.2xu,-0.5xu) and (1.2xu,-0.5xu) .. (1.2xu,0xu)) shifted (0xu,0.05xu) pensize (1.0pt) withcolor bcolor;
	midarrowblue ((-1.2xu,0.05xu) ..controls (-1.2xu,-0.45xu) and (1.2xu,-0.45xu) .. (1.2xu,0.05xu),0.6);
	draw (-1xu,0.03xu){dir -20}..(1xu,0.03xu) pensize (1.0pt);
	draw (-0.8xu,-0.02xu){dir 25}..(0.8xu,-0.02xu) pensize (1.0pt);
	draw (0xu,-0.15xu) .. controls (-0.1xu,-0.18xu) and (-0.1xu,-0.92xu) .. ( 0xu,-0.95xu) pensize (1.0pt) withcolor ecolor;
	midarrowred ((0xu,-0.15xu) .. controls (-0.1xu,-0.18xu) and (-0.1xu,-0.92xu) .. ( 0xu,-0.95xu),0.65);
	
	draw halfcircle xscaled 1.9xu yscaled 3.5xu rotated -90 ..(-1.5xu,0.7xu){left}..(-3xu,0.95xu){left} pensize (1pt);
	draw halfcircle xscaled 1.9xu yscaled 3.5xu rotated 90 shifted (-3xu,0xu) ..(-1.5xu,-0.7xu){right}..(0xu,-0.95xu){right} pensize (1pt);
	
	label(btex $\mathfrak{A}_1$ etex, (-2.7xu, -1.2xu));
	label(btex $\mathfrak{A}_2$ etex, (0.3xu, -1.2xu));
	label(btex $\mathfrak{B}_1$ etex, (-1.63xu, 0.3xu));
	label(btex $\mathfrak{B}_2$ etex, (1.37xu, 0.3xu));
	
	dotlabeldiam:=4pt;
	dotlabel.bot (btex $ $ etex, (-1.7xu,-0.4xu));
	path cpath; cpath = fullcircle xscaled 0.47xu yscaled 0.38xu rotated (10) shifted (-1.7xu,-0.4xu);
	draw cpath withcolor gcolor pensize (1pt);
	midarrowgreen(cpath,0.8);
	label.rt(btex $\mathfrak{C}$ etex, point 0.4 of cpath);
	
\end{mpostfig}

\begin{mpostfig}[label=highergenus]

draw ((-3xu,-0.143xu) .. controls (-2.9xu,-0.18xu) and (-2.9xu,-0.90xu) .. ( -3xu,-0.915xu)) shifted (0xu,-0.3xu) rotated -5 dashed evenly pensize (1.0pt) withcolor ecolor;
draw ((-4xu,0xu) ..controls (-4xu,0.5xu) and (-1.6xu,0.5xu) .. (-1.6xu,0xu)) shifted (-0.3xu,-0.25xu) rotated -5 pensize (1pt) withcolor bcolor;
draw ((-4xu,0xu) ..controls (-4xu,-0.5xu) and (-1.6xu,-0.5xu) .. (-1.6xu,0xu)) shifted (-0.3xu,-0.25xu) rotated -5 pensize (1.0pt) withcolor bcolor;
midarrowblue (((-4xu,0.05xu) ..controls (-4xu,-0.45xu) and (-1.6xu,-0.45xu) .. (-1.6xu,0.05xu)) shifted (-0.3xu,-0.3xu) rotated -5,0.6);
draw ((-4xu,0.03xu){dir -20}..(-2xu,0.03xu)) shifted (-0.1xu,-0.3xu) rotated -5 pensize (1.0pt);
draw ((-3.8xu,-0.02xu){dir 25}..(-2.2xu,-0.02xu)) shifted (-0.1xu,-0.3xu) rotated -5 pensize (1.0pt);
midarrowred (((-3xu,-0.143xu) .. controls (-3.1xu,-0.18xu) and (-3.1xu,-0.90xu) .. (-3xu,-0.915xu)) shifted (0xu,-0.3xu) rotated -5,0.65);

draw ((0xu,-0.153xu) .. controls (0.1xu,-0.18xu) and (0.1xu,-0.72xu) .. ( 0xu,-0.74xu)) shifted (0.1xu,-0.4xu) rotated -10 dashed evenly pensize (1.0pt) withcolor ecolor;
draw ((-1.xu,0xu) ..controls (-1.0xu,0.5xu) and (1.0xu,0.5xu) .. (1.0xu,0xu)) shifted (0.1xu,-0.4xu) rotated -7 pensize (1.0pt) withcolor bcolor;
draw ((-1.xu,0xu) ..controls (-1.xu,-0.5xu) and (1.xu,-0.5xu) .. (1.xu,0xu)) shifted (0.1xu,-0.4xu) rotated -7 pensize (1.0pt) withcolor bcolor;
midarrowblue (((-1.xu,0xu) ..controls (-1.xu,-0.5xu) and (1.xu,-0.5xu) .. (1.xu,0xu)) shifted (0.1xu,-0.4xu) rotated -7,0.6);
draw ((-0.7xu,0.03xu){dir -30}..(0.7xu,0.03xu)) shifted (0.1xu,-0.4xu) rotated -7 pensize (1.0pt);
draw ((-0.6xu,-0.02xu){dir 25}..(0.6xu,-0.02xu)) shifted (0.1xu,-0.4xu) rotated -7 pensize (1.0pt);
midarrowred (((0xu,-0.153xu) .. controls (-0.1xu,-0.18xu) and (-0.1xu,-0.72xu) .. ( 0xu,-0.74xu)) shifted (0.1xu,-0.4xu) rotated -10,0.65);

draw ((0xu,-0.48xu) .. controls (0.15xu,-0.5xu) and (0.15xu,-1.18xu) .. ( 0xu,-1.2xu)) shifted (3.1xu,-.9xu) rotated 30 dashed evenly pensize (1.0pt) withcolor ecolor;
draw ((-1.1xu,0xu) ..controls (-1.1xu,0.6xu) and (1.1xu,0.6xu) .. (1.1xu,0xu)) shifted (3.1xu,-.9xu) rotated 25 pensize (1.0pt) withcolor bcolor;
draw ((-1.1xu,0xu) ..controls (-1.1xu,-0.6xu) and (1.1xu,-0.6xu) .. (1.1xu,0xu)) shifted (3.1xu,-.9xu) rotated 25 pensize (1.0pt) withcolor bcolor;
midarrowblue (((-1.1xu,0xu) ..controls (-1.1xu,-0.6xu) and (1.1xu,-0.6xu) .. (1.1xu,0xu)) shifted (3.1xu,-.9xu) rotated 25,0.65);
draw ((-0.9xu,0.03xu){dir -30}..(0.9xu,0.03xu)) shifted (3.1xu,-.9xu) rotated 25 pensize (1.0pt);
draw ((-0.8xu,-0.02xu){dir 25}..(0.8xu,-0.02xu)) shifted (3.1xu,-.9xu) rotated 25 pensize (1.0pt);
midarrowred (((0xu,-0.48xu) .. controls (-0.15xu,-0.5xu) and (-0.15xu,-1.18xu) .. ( 0xu,-1.2xu)) shifted (3.1xu,-.9xu) rotated 30,0.65);

draw halfcircle xscaled 1.8xu yscaled 3.5xu rotated 90 shifted (-3xu,-0.05xu) ..(-1.5xu,-0.7xu){right}..(0.3xu,-1.2xu){right}..(1.1xu,-1.0xu){dir 20} pensize (1pt);
draw (1.1xu,-1.0xu){dir 20}..(2.1xu,-0.5xu){right} dashed evenly pensize (1pt);
draw (2.1xu,-0.5xu){right}..(2.9xu,-0.55xu){right}..(3.6xu,-0.36xu){dir 30} pensize (1pt);
draw (-3xu,0.85xu){right}..(-1.0xu,0.3xu){right}..(0.xu,0.3xu){right}..(1.1xu,0.1xu){right} pensize (1pt);
draw (1.1xu,0.1xu){right}..(2.1xu,0.6xu){dir 40} dashed evenly pensize (1pt);
draw (2.1xu,0.6xu){dir 40}..(2.7xu,1.2xu){dir 30} pensize (1pt);
draw halfcircle xscaled 1.8xu yscaled 3.xu rotated 300 shifted (3.15xu,0.42xu) pensize (1pt);

draw ((-3xu,-0.143xu) .. controls (-3.1xu,-0.18xu) and (-3.1xu,-0.90xu) .. (-3xu,-0.915xu)) shifted (0xu,-0.3xu) rotated -5 pensize (1.0pt) withcolor ecolor;
draw ((0xu,-0.153xu) .. controls (-0.1xu,-0.18xu) and (-0.1xu,-0.72xu) .. ( 0xu,-0.74xu)) shifted (0.1xu,-0.4xu) rotated -10 pensize (1.0pt) withcolor ecolor;
draw ((0xu,-0.48xu) .. controls (-0.15xu,-0.5xu) and (-0.15xu,-1.18xu) .. ( 0xu,-1.2xu)) shifted (3.1xu,-.9xu) rotated 30 pensize (1.0pt) withcolor ecolor;

label(btex $A_1$ etex, (-2.9xu, -1.2xu)) withcolor ecolor;
label(btex $A_2$ etex, (-0.3xu, -1.3xu)) withcolor ecolor;
label(btex $A_h$ etex, (4.03xu, -0.43xu)) withcolor ecolor;
label(btex $B_1$ etex, (-2.8xu, 0.57xu)) withcolor bcolor;
label(btex $B_2$ etex, (1.3xu, -0.3xu)) withcolor bcolor;
label(btex $B_h$ etex, (2.8xu, -0.35xu)) withcolor bcolor;
label(btex \Large$\Sigma^\times$ etex, (0.9xu, 0.65xu));

dotlabeldiam:=4pt;
dotlabel.bot (btex $ $ etex, (-1.55xu,-0.2xu));
path cpath; cpath = fullcircle xscaled 0.47xu yscaled 0.38xu rotated (10) shifted (-1.55xu,-0.2xu);
draw cpath withcolor gcolor pensize (1pt);
midarrowgreen(cpath,0.8);
label.rt(btex $C$ etex, point 0.4 of cpath) withcolor gcolor;

\end{mpostfig}

\usepackage{graphicx}
\usepackage{subcaption}
\usepackage{caption}

\usepackage{ragged2e}

\usepackage[dvipsnames,table]{xcolor}
\definecolor{mygreen}{rgb}{0,0.4,0}
\definecolor{myblue}{rgb}{0,0.0,0.4}
\definecolor{refrcolor}{rgb}{0,0.4,0}
\definecolor{cgreen}{rgb}{0,0.7,0}
\definecolor{ecolor}{rgb}{.52,.03,.06}
\definecolor{bgcolor}{rgb}{.96,.95,.80}
\definecolor{bgcolordark}{rgb}{.80,.80,.67}
\definecolor{faint}{rgb}{.80,.80,.80}

\usepackage{tcolorbox}
\tcbuselibrary{breakable, skins}
\usetikzlibrary{patterns}

\newtcolorbox{myproofbox}[1][]{
	enhanced,
	breakable,
	borderline west={3pt}{0pt}{faint},
	notitle,
	before skip=10pt,
	after skip=10pt,
	colback=white, 
	colframe=white,
	frame hidden,
	boxrule=0pt, 
	boxsep=0pt,
	sharp corners,
	left=9pt, right=0pt, top=1pt, bottom=0pt,
	fontupper=\small,
	#1
}

\makeatletter
	{\popQED\endtrivlist\end{myproofbox}\@endpefalse%
	\if@noskipsec\leavevmode\fi\noindent\ignorespacesafterend}
\makeatother

\newtheorem{example}[theorem]{Example}

\newtcolorbox{myexamplebox}[1][]{%
	enhanced,
	breakable,
	borderline west={3pt}{0pt}{faint},
	notitle,
	before skip=10pt,
	after skip=10pt,
	colback=white, 
	colframe=white,
	frame hidden,
	boxrule=0pt, 
	boxsep=0pt,
	sharp corners,
	left=9pt, right=0pt, top=1pt, bottom=0pt,
	#1
}

\makeatletter
	{\endtrivlist\end{myexamplebox}\@endpefalse%
	\if@noskipsec\leavevmode\fi\noindent\ignorespacesafterend}
\makeatother

\usepackage{enumitem}

\usepackage{tabularx}
\usepackage{booktabs}

\makeatletter
\providecommand*{\shuffle}{%
	\mathbin{\mathpalette\shuffle@{}}%
}
\newcommand*{\shuffle@}[2]{%
	\sbox0{$#1\vcenter{}$}%
	\kern .15\ht0 
	\rlap{\vrule height .25\ht0 depth 0pt width 2.5\ht0}%
	\raise.1\ht0\hbox to 2.5\ht0{%
		\vrule height 1.75\ht0 depth -.1\ht0 width .17\ht0 %
		\hfill
		\vrule height 1.75\ht0 depth -.1\ht0 width .17\ht0 %
		\hfill
		\vrule height 1.75\ht0 depth -.1\ht0 width .17\ht0 %
	}%
	\kern .15\ht0 
}
\makeatother

\usepackage[pdfencoding=auto,bookmarks=true,hyperfigures=true]{hyperref}%
\PassOptionsToPackage{unicode}{hyperref}
\providecommand{\hypersetup}[1]{}

\hypersetup{plainpages=false}
\hypersetup{pdfpagemode=UseNone}
\hypersetup{bookmarksnumbered=true}
\hypersetup{pdfstartview=FitH}
\hypersetup{colorlinks=false}
\hypersetup{citebordercolor={0.7 0.7 1}}
\hypersetup{urlbordercolor={.4 .8 1}}
\hypersetup{linkbordercolor={1 .8 .6}}
\hypersetup{colorlinks=true, urlcolor=[rgb]{0.13,0.30,0.45}, linkcolor=[rgb]{0.13,0.30,0.45}, citecolor=[rgb]{0.55,0.0,0.05}}

\usepackage{tocloft}

\makeatletter

\newcommand{\l@appendixsec}[2]{%
	\vspace{0.1ex}%
	\@dottedtocline{1}{\cftsubsecindent}{\cftsubsecnumwidth}{#1}{#2}%
}
\newcommand{\l@appendixsubsec}[2]{%
	\vspace{0.1ex}%
	\@dottedtocline{2}{\cftsubsubsecindent}{\cftsubsubsecnumwidth}{#1}{#2}%
}
\providecommand*{\toclevel@appendixsec}{1}
\providecommand*{\toclevel@appendixsubsec}{2}
\def\Hy@toc@appendixsec{section}
\def\Hy@toc@appendixsec{subsection}
\makeatother

\newif\ifpublic\publicfalse
\publictrue

\newif\ifnote\notetrue
\ifpublic
\notefalse
\fi

\let\Re\relax\DeclareMathOperator{\Re}{Re}
\let\Im\relax\DeclareMathOperator{\Im}{Im}

\let\Re\undefined\DeclareMathOperator{\Re}{Re}
\let\Im\undefined\DeclareMathOperator{\Im}{Im}

\DeclareMathOperator{\Ad}{Ad}
\DeclareMathOperator{\ad}{ad}

\DeclareMathOperator{\Res}{Res}

\ProvideMathFix{rfrac,vfrac}
\newcommand{\iunit}{{\mathring{\imath}}}

\newcommand{\der}{\mathrm{d}}
\newcommand{\diff}[2][.]{\mathinner{\der#2\if #1.\else^{#1}\fi}}

\newcommand{\alg}[1]{\mathfrak{#1}}

\newcommand{\grpdot}{{\cdot}}

\let\qed\relax\newcommand{\qed}
{\hfill\ensuremath{\Box}}

\newcommand{\pd}{\partial}

\newcommand{\dd}{\mathrm{d}}

\newcommand{\cF}{\mathcal{F}}       
\newcommand{\cG}{\mathcal{G}} 
\newcommand{\cH}{\mathcal{H}}

\newcommand{\cK}{\mathcal{K}}
\newcommand{\cL}{\mathcal{L}}

\newcommand{\cO}{\mathcal{O}}

\newcommand{\cU}{\mathcal{U}}

\newcommand{\genus}{\ensuremath{h}}

\newcommand{\RSurf}{\Sigma}

\newcommand{\texteq}{\,{=}\,}

\ExplSyntaxOn
\NewDocumentCommand{\args}{o o}
{
	\IfValueTF{#1}
	{
		\IfValueTF{#2}
		{
			\seq_set_from_clist:Nn \l_tmpa_seq { #1 }
			\left( \seq_use:Nn \l_tmpa_seq { ,~ } ~;~ #2\right)
		}{
			\seq_set_from_clist:Nn \l_tmpa_seq { #1 }
			\left( \seq_use:Nn \l_tmpa_seq { ,~ }\right)
		}
	}{
		\IfValueTF{#2}
		{
			\left( #2 \right)
		}{
			
		}
	}
}
\ExplSyntaxOff

\newcommand{\hd}{\omega}

\newcommand{\cdhs}[2]{\cK_{\mathsf{DHS}}(#1,p;#2)}
\newcommand{\cenr}[2]{\cK_\mathsf{E}(#1,p;#2)}
\DeclareMathOperator{\genf}{\mathrm{L}}
\DeclareMathOperator{\svgenf}{\cL}
\newcommand{\genfE}[2]{\mathrm{L}_\mathsf{E}(#1,p;#2)}
\newcommand{\svgenfE}[2]{\cL(#1,p;#2)}
\newcommand{\genfDHS}[2]{\mathrm{L}_\mathsf{DHS}(#1,p;#2)}
\newcommand{\gauge}[2]{g(#1,p;#2)}
\newcommand{\monoA}[2]{\mathrm{A}_{#1}(#2)}
\newcommand{\monoB}[2]{\mathrm{B}_{#1}(#2)}
\newcommand{\bfunc}{\mathrm{T}}

\newcommand{\SVaut}{\mathsf{\sigma}}
\newcommand{\DHSaut}{\mathsf{\xi}}

\newcommand{\pRSurf}{\RSurf^\times}
\newcommand{\RSurfuniv}{\RSurf_\mathrm{univ}}
\newcommand{\pRSurfuniv}{\pRSurf_\mathrm{univ}}
\newcommand{\fg}{\pi_1(\RSurf)}
\newcommand{\fgp}{\pi_1(\pRSurf)}
\newcommand{\univpt}[1]{#1}
\newcommand{\univppt}[1]{#1}

\newcommand{\Id}{\mathbb{1}}
\newcommand{\iso}{\cong}

\newcommand{\free}{\alg{u}}
\newcommand{\uea}{\cU}
\newcommand{\diffs}[2]{\Omega_{#1}(#2)}
\newcommand{\funcs}[1]{\cO(#1)}

\newcommand{\ant}{P}
\newcommand{\ara}[2]{\mathcal{G}(#1,#2)}

\usepackage{bbold} 
\usepackage{lipsum}
\usepackage{tikz-cd} 

\begin{document}


\title{Single-valued polylogarithms for higher genera}

\author{Konstantin Baune${}^\ast$}
\author{Johannes Broedel${}^\ast$}
\author{Yannis Moeckli}
\thanks{Electronic adresses: \texttt{\{baunek|jbroedel|moeckliy\}@ethz.ch}}
\affiliation{%
 Institute for Theoretical Physics, ETH Zürich\\Wolfgang-Pauli-Str.~27, 8093 Zurich, Switzerland
}%

\date{\today}

\begin{abstract}
	\noindent We extend the construction of single-valued polylogarithms at genus one from \href{https://arxiv.org/abs/2511.15240}{arXiv:2511.15240} to once-punctured Riemann surfaces of higher genera. The resulting functions have a trivial monodromy representation with respect to the fundamental group, hence they descend to well-defined functions on the surface.
	Our construction of single-valued polylogarithms is based on Enriquez' connection and relates them to the polylogarithms from D'Hoker--Hidding--Schlotterer. Finally, we identify the Arakelov Green's function within our framework.
\end{abstract}


\maketitle

\section{\label{sec:intro}Introduction}

\vspace{0.025cm}

Efficient calculations of scattering amplitudes in QFT and string theory rely on a profound understanding of the underlying function space. 
For most situations, these spaces can be chosen to be spanned by polylogarithms, various flavors of which have have been studied extensively in e.g.~\rcites{Goncharov:1998kja,Goncharov:2001iea,BrownThesis,BrownLevin,CEE,DHoker:2023vax,EZ1,EZ2,Baune:2024biq,Broedel:2017kkb,EZ3,Schlotterer:2025qjv,Baune:2025ndr,DHoker:2025szl,DHoker:2025dhv,DHoker:2024ozn,Baune:2024ber,DHoker:2026ggx,DHoker:2026lgg}.

While for many QFT calculations it suffices to use genus-zero polylogarithms~\cite{Travaglini_2022,PanzerPhD}, certain loop integrals are described by elliptic polylogarithms~\cite{Broedel:2017kkb,Broedel:2017siw,Bourjaily:2022bwx} or generalizations to higher genera \cite{Marzucca:2023gto,Duhr:2024uid}.
In general, polylogarithms are multi-valued functions; certain classes of QFT scattering amplitudes, however, evaluate to single-valued combinations of genus-zero polylogarithms~\cite{BrownSVMPL,Dixon:2012yy}.

In string theory, the understanding of genus-zero and elliptic polylogarithms has proven useful in the calculation of tree-level~\cite{Broedel:2013aza,Broedel:2013tta,Brown:2019wna} and one-loop~\cite{Broedel:2014vla,Broedel:2017jdo,Broedel:2018izr,Broedel:2019gba,Broedel:2020tmd} open-string amplitudes, respectively, while it is expected that \emph{higher-genus polylogarithms (hgPLs)} \cite{EnriquezHigher,EZ1,EZ2,EZ3,DHoker:2023vax,Baune:2024biq,DHoker:2024ozn,Baune:2024ber,DHoker:2025szl,DHoker:2025dhv,DHoker:2025jgb,Baune:2025sfy} describe higher-loop open-string amplitudes. 
On the other hand, single-valued polylogarithms are a suitable class of functions for describing closed-string scattering at genus zero and genus one~\cite{Baune:2025ndr,Schlotterer:2025qjv,Stieberger:2013wea,Stieberger:2022lss,Stieberger:2014hba,Brown:2019wna,Brown:2018omk}. Following the open-string logic, it is expected that \emph{single-valued higher-genus polylogarithms (svhgPLs)} are the correct class of functions for closed-string scattering at higher loop orders. 

In this letter, we generalize the genus-one construction of svhgPLs in \rcite{Baune:2025ndr} to Riemann surfaces of higher genera. 


\vfill

\section{Prerequisites}\label{sec:rev_rstheory}

In the following, we review several mathematical concepts.
Let $\RSurf$ be a compact Riemann surface of genus $\genus\,{\geq}\,1$, $p\,{\in}\,\RSurf$ an arbitrary point and $\pRSurf\,{\coloneqq}\,\RSurf\,{\setminus}\,\{p\}$ the punctured Riemann surface.
To describe multi-valued objects, we fix universal covers $\pi\,{:}\,\RSurfuniv\,{\rightarrow}\,\RSurf$ and $\pi^\times\,{:}\,\pRSurfuniv\,{\rightarrow}\,\pRSurf$ and also define $\RSurfuniv^p\,{\coloneqq}\,\RSurfuniv\,{\setminus}\,\pi^{-1}(p)$.
By slight abuse of notation, we will denote points on the surface as well as on the universal covers by the same letters.

The fundamental group of $\pRSurf$ can be presented as \cite{Hatcher2002}
\begin{equation}
	\begin{aligned}
		\fgp &= \bigg\langle A_i,B_i,C\,\Big|\,i\in[\genus],\,\prod_{j=h}^1 B^{-1}_jA^{-1}_jB_jA_j=C \bigg\rangle ,\label{eqn:presfg}
	\end{aligned}
\end{equation}
where $[\genus]\,{\coloneqq}\,\{1,\ldots,\genus\}$. The generators $A_i,B_i$ correspond to the $2\genus$ non-contractible loops around the handles of the surface, whereas $C$ describes a loop around the puncture $p$:
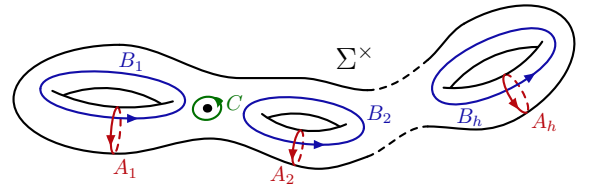
\begin{equation*}
		\mpostuse[scale=0.8]{highergenus}
\end{equation*}
The fundamental group of $\RSurf$ can then be obtained as $\fg\,{\iso}\,\fgp/\langle C\rangle$.
For a point $\univpt{z}\,{\in}\,\RSurfuniv$ (resp.~$\univppt{z}\,{\in}\,\pRSurfuniv$), we denote by $\gamma\grpdot \univpt{z}$ for $\gamma\,{\in}\,\fg$ (resp.~$\gamma\,{\in}\,\fgp$) the endpoint of the unique lift of $\gamma$ starting at $z$. This naturally extends to an action $\gamma\,{\mapsto}\,\gamma\grpdot z$ of the fundamental groups $\fg$ and $\fgp$ on the universal covers $\RSurfuniv$ and $\pRSurfuniv$ respectively. By construction, these actions preserve the fibers $\pi^{-1}(z)$ (resp.~$\pi_\times^{-1}(z)$) of the covering maps, hence the action of $\fg$ restricts to an action on $\RSurfuniv^p$ (see e.g.~\rcite{EZ1,Hatcher2002} for details).

Having described the topological and geometric setting, let us briefly introduce several algebraic notions. For $X\,{\coloneqq}\,\{a_i,b_i|i\,{\in}\,[\genus]\}$ we denote by $X^*\,{\subset}\,\uea$ the set of all words generated by $X$, by $\free$ (the degree-completion of) the Lie algebra freely generated by the set $X$ and by $\uea$ its (completed) universal enveloping algebra. Furthermore, we regard $\free\subset\uea$ via its natural inclusion.

$\uea$ naturally carries the structure of a Hopf algebra w.r.t.~concatenation, with the coproduct given by declaring the generators $X$ to be primitive and the \hypertarget{target:antipode}antipode $\ant:\uea\,{\rightarrow}\,\uea$ to be the unique anti-homomorphism such that $\ant(a_i)\texteq{-}\,a_i$ and $\ant(b_i)\texteq{-}\,b_i$ for all $i\,{\in}\,[\genus]$ \cite{Brownhyperlogs}. We denote its group of group-like elements by $\cG\cU$. 

Below, we will use a shorthand notation for automorphisms of $\uea$: since such a map $\varrho:\uea\,{\rightarrow}\,\uea$ is entirely determined by its image of the set $X$ of generators, we conveniently write $X\,{\mapsto}\, X_\varrho\,{\coloneqq}\,\{\varrho(a_i),\varrho(b_i)|i\,{\in}\,[\genus]\}$. Moreover, we will write $S(X)$ for an element of $\uea$, thereby implicitly regarding it as an expansion in the generators $X$. The automorphism $\varrho$ will then be denoted by $S(X_\varrho)\,{\coloneqq}\, \varrho(S(X))$. 

Finally, the choice of generators $\{A_i,B_i,C|i\,{\in}\,[\genus]\}$ of $\fgp$ is such that their images in the first homology group $H_1(\pRSurf)\,{\iso}\,\fgp/[\fgp,\fgp]$ define a symplectic basis w.r.t.~the intersection pairing \cite{Bobenko:2011}. Denoting the corresponding dual basis of holomorphic differentials by $\hd_j$, $j\,{\in}\,[\genus]$, the period matrix reads
\begin{equation}
	\tau_{ij} = \int_{B_i}\hd_j \, , \qquad i,j\in[\genus] \, ,
\end{equation}
where we conveniently used the same symbol for the generator in $\fgp$ and its image in $H_1(\pRSurf)$.
All objects defined in the remainder of this letter will depend on the geometry of the surface, captured by $\tau_{ij}$, but we will omit this dependence throughout this letter as we regard the geometry as fixed.

Moreover, for $M$ a smooth manifold and $D\,{\subset}\, M$ a discrete subset, we generally denote the space of (smooth) functions on $M$ and differential (one-)forms on $M$ with at most simple poles on $D$ as $\funcs{M}$ and $\diffs{D}{M}$ respectively. Conventionally, we will omit the subscript $D$ if $D\texteq\varnothing$ as well as write $\Omega_x(M)$ if $D\texteq\{x\}$ for $x\,{\in}\, M$.

\section{\label{sec:review}Constructing polylogarithms}

\noindent\textit{\textbf{Flat connections.}} Assume a flat\footnote{The connection is said to be flat if $\dd\cK(z,p;X)-\cK(z,p;X)\,{\wedge}\,\cK(z,p;X)\texteq0$ for all $z\in\pRSurf$.} connection on the manifold $\pRSurf$, which is valued in the Lie algebra~$\free$. We suppose that this connection can be written as $\dd - \cK(z,p;X)$ for $\cK\in\diffs{\pi^{-1}(p)}{\RSurfuniv}\,{\otimes}\,\free$ obeying
\begin{align}
	\cK(\gamma\grpdot z,p;X) = \Ad(\phi(\gamma))\,\cK(z,p;X) \, ,
\end{align}
where $\phi:\fg\rightarrow\cG\uea$ denotes a group homomorphism.

\smallskip
\noindent\textit{\textbf{Polylogarithms.}} In this letter, we understand polylogarithms $\genf(z,p;X)$ (w.r.t.~$\cK(z,p;X)$) as a solution to the differential equation 
\begin{align}\label{eqn:KZB-general}
	(\dd-\cK(z,p;X))\genf(z,p;X)=0\, ,
\end{align}
which takes values in $\funcs{\pRSurfuniv}\,{\otimes}\,\uea$ and is subject to a certain normalization condition. 
 Since \eqn{eqn:KZB-general} is a linear first-order differential equation, it follows that any two invertible solutions differ by at most a constant element $\mathrm{C}$ of $\uea$, that is
 \begin{equation}\label{eqn:uniqueness}
	\genf'(\univppt{z},p;X) = \genf(z,p;X)\,\mathrm{C}
\end{equation}
for any other solution $\mathrm{L}'(z,p;X)$ and $\mathrm{C}$ independent of $z\,{\in}\,\pRSurfuniv$.
Finally, in the scope of this letter, we exclusively consider solutions $\genf(z,p;X)$ that are group-like, i.e.~elements of $\cG\uea$ for fixed $z\,{\in}\,\pRSurfuniv$. In this way, the coefficients of $\genf(z,p;X)$ determined by
\begin{equation}\label{eqn:sol}
	\genf(z,p;X) = \sum_{x\in X^*}\ell_x(\univppt{z},p)\,x
\end{equation}
upon expansion in the set of generators $X$ automatically obey the shuffle relations~\cite{Brownhyperlogs}. Restricting to group-like solutions also implies the constants $\mathrm{C}$ in \eqn{eqn:uniqueness} are elements of $\cG\uea$.

\smallskip
\noindent\textit{\textbf{Monodromy representation.}} Polylogarithms will generically admit non-trivial monodromies when analytically continuing them around a non-contractible cycle of $\pRSurf$. Monodromies of a solution $\genf(z,p;X)$ can be conveniently described in terms of the action of $\fgp$ on $\pRSurfuniv$. In particular, one can check that the function $\phi(\gamma)\genf(\gamma\grpdot z,p;X)$ again constitutes a solution to the differential equation~\eqref{eqn:KZB-general}. Therefore, by the uniqueness statement~\eqref{eqn:uniqueness}, we can write
\begin{equation}\label{eqn:monodromy_rep}
	\genf(\gamma\grpdot z,p;X) = \phi(\gamma^{-1})\genf(z,p;X)\,M_{\genf}(\gamma),
\end{equation} 
for $M_{\genf}\,{:}\,\fgp\,{\rightarrow}\,\cG\uea$ a group homomorphism. $M_{\genf}$ is referred to as the \emph{monodromy representation} of the solution $\genf(z,p;X)$. Monodromy representations in the context of polylogarithms on Riemann surfaces of genus zero and one have been explored in \rcites{Brownhyperlogs,Brown:2018omk} and \rcites{Baune:2025ndr,Schlotterer:2025qjv}, respectively.

\vfill

\section{\lowercase{svhg}PL\lowercase{s}: general mechanism}

As indicated above, we aim to construct a class of single-valued polylogarithms, i.e.~a set of functions with trivial monodromy representation, whose properties resemble those of their multi-valued cousins as closely as possible. In particular shuffle relations should be preserved. In \rcite{Brownhyperlogs}, such a class has been described by Brown for the punctured Riemann sphere, and it has been generalized to elliptic curves in \rcites{Schlotterer:2025qjv,Baune:2025ndr}.

Within our framework, the echo of Brown's procedure is as follows: assuming a non-trivial automorphism\footnote{The defining property of $\iota$ in fact implies that it acts as an involution on $M_{\genf}(\fgp)\,{\subset}\,\cG\uea$.} $\iota:\cG\uea\,{\rightarrow}\,\cG\uea$ such that $\iota\,{\circ}\, M_{\genf} \texteq M_{\genf}$, we can define the function
\begin{equation}
	\widetilde\svgenf(z,p;X) \coloneqq \iota\big(\genf(z,p;X)\big)\genf(z,p;X)^{-1} \, ,
\end{equation}
for which \eqn{eqn:monodromy_rep} implies
\begin{equation}
	\begin{aligned}\label{eqn:almostSV}
		\widetilde\svgenf(\gamma\grpdot z,p;X) &= \iota(\genf(\gamma\grpdot z,p;X))\genf(\gamma\grpdot z,p;X)^{-1} \\
		&= \iota(\phi(\gamma^{-1}))\,\widetilde\svgenf(z,p;X)\,\phi(\gamma) \, .
	\end{aligned}
\end{equation}
Given the above equation, $\widetilde\svgenf(z,p;X)$ is not quite single-valued yet. \hypertarget{target:T} The remnants can be cancelled by assuming the existence of a suitable (multi-valued) function $\mathrm{T}(z,p;X)\in\funcs{\RSurfuniv}\otimes\cG\cU$ such that $M_\mathrm{T}\texteq\phi$. We call such a function $\bfunc(z,p;X)$ a \emph{trivialization}. Accordingly,
\begin{equation}\label{eqn:ansatz-general}
	\svgenf(z,p;X) \coloneqq \iota(\mathrm{T}(z,p;X))\,\widetilde\svgenf(z,p;X)\,\mathrm{T}(z,p;X)^{-1}
\end{equation}
is a single-valued function, i.e.~it is well-defined on $\pRSurf$. Since all constituents are group-like, the coefficients of $\svgenf(z,p;X)$ adhere to the shuffle relations.

\section{\lowercase{svhg}PL\lowercase{s}: explicit construction}\label{sec:rev_hgmpls}

In this section, we will show by explicit construction that such a trivialization $\mathrm{T}(z,p;X)$ as well as a map $\iota$ exist for the connection defined by Enriquez in \rcite{EnriquezHigher}. In a second step, we relate the construction to the connection constructed by D'Hoker--Hidding--Schlotterer~\cite{DHoker:2023vax}. Finally, we will discuss the relation to the elliptic scenario discussed in \rcite{Baune:2025ndr}.

\smallskip
\noindent\textit{\textbf{svhgPLs from Enriquez' connection.}}\label{sec:svhgplE} We consider the differential form $\cenr{\univpt{z}}{X}\,{\in}\,\diffs{\pi^{-1}(p)}{\RSurfuniv}\,{\otimes}\,\free$ originating from a flat connection constructed in \rcite{EnriquezHigher}, which is characterized by the properties
\begin{subequations}\label{eqn:KEprops}
    \begin{align}
        \cenr{A_j\grpdot \univpt{z}}{X} &= \cenr{\univpt{z}}{X} \, , \\
        \label{eqn:condB}\cenr{B_j\grpdot \univpt{z}}{X} &= e^{-2\pi\iunit b_j}\cenr{\univpt{z}}{X}\,e^{2\pi\iunit b_j} \, , \\
        \hspace{-0.5ex}\label{eqn:condRes}\Res_{\univpt{z}= p}\cenr{\univpt{z}}{X} &= \sum_{k=1}^\genus \ad(b_k)\,a_k \eqqcolon t \, ,
    \end{align}
\end{subequations}
where $j\,{\in}\,[\genus]$. Notice that \eqn{eqn:KEprops} implies that $\cenr{\univpt{z}}{X}$ admits poles at all translates of $p$, i.e.~at all points in $\pi^{-1}(p)\,{\subset}\,\RSurfuniv$.
$\cenr{z}{X}$ gives rise to a generalization of the elliptic KZB equation~\cite{CEE,Matthes:Thesis} to higher genera. Specializing \eqn{eqn:KZB-general}, we are going to consider the differential equation
\begin{equation}
	\label{eqn:E-KZB}
	\dd \genfE{\univppt{z}}{X} = \cenr{\univppt{z}}{X}\,\genfE{\univppt{z}}{X} \, ,
\end{equation} 
whose solution will naturally be an element of $\funcs{\pRSurfuniv}\otimes\uea$ due to the simple poles of $\cenr{\univpt{z}}{X}$. It can be shown that a particular solution $\genfE{\univppt{z}}{X}$ can be constructed in terms of iterated integrals (w.r.t.~a tangential basepoint\footnote{Tangential basepoint regularization for these hgPLs has been formulated in \cite{Baune:2025sfy} using Schottky uniformization and is compatible with shuffle relations. In the following, we will assume the existence of a valid solution with the desired properties.} in $\pRSurfuniv$) of $\cenr{z}{X}$. It obeys the asymptotics
\begin{equation}\label{eqn:asymptotics}
	\genfE{\univppt{z}}{X}\,e^{-t\log(\univppt{z})}\overset{z\rightarrow p}{\sim}1 \, ,
\end{equation}
which unpacks to
\begin{equation}
	\genfE{v}{X} = f(v)\,e^{t\log(v)}
\end{equation}
in a local coordinate $v$ such that $v(p)=0$, whereas $f$ is a (locally) holomorphic function satisfying $f(0)=1$. Being (regularized) iterated integrals, the coefficients of $\genfE{z}{X}$ satisfy the shuffle relations. The latter in particular implies that $\genfE{z}{X}$ is group-like. Finally, the solution expands to lowest order as
\begin{equation}
	\label{eqn:exp_loworder}
	\genfE{z}{X} = 1 + \sum_{j=1}^\genus\int_p^z\omega_j(z)\,a_j + \ldots \, .
\end{equation}
We now want to use the (multi-valued) $\genfE{z}{X}$ in order to construct a family of single-valued functions on $\pRSurf$ along the lines discussed in the previous section. To do this, we will generalize the constructions of refs.~\cite{Brownhyperlogs, Baune:2025ndr} to the topological setting at hand.  
We begin by defining the \emph{monodromy series} $\monoA{i}{X},\monoB{i}{X}\,{\in}\,\cG\uea$ for $i\,{\in}\,[\genus]$ as
\begin{subequations}\label{eqns:hgPTOs}
	\begin{align}
		\label{eqn:EmonoA}\monoA{i}{X} &\coloneqq \genfE{\univppt{z}}{X}^{-1}\,\genfE{A_i\grpdot \univppt{z}}{X}, \\
		\label{eqn:EmonoB}\monoB{i}{X} &\coloneqq \genfE{\univppt{z}}{X}^{-1}\,e^{2\pi\iunit b_i}\,\genfE{B_i\grpdot \univppt{z}}{X} .
	\end{align}
\end{subequations}
It follows from the uniqueness statement~\eqref{eqn:uniqueness} that both of these are indeed independent of $\univppt{z}\,{\in}\,\pRSurfuniv$. Also, the coefficients of the monodromy series $\monoA{i}{X}$ can be related to the higher-genus multiple zeta values studied in \rcite{Baune:2025sfy}.

The definition of the monodromy series, together with the presentation of $\fgp$ given in \eqn{eqn:presfg}, implies
\begin{equation}
\begin{aligned}\label{eqn:CcycleAction}
	&\genfE{C\grpdot\univppt{z}}{X} = \mathrm{L}_\mathrm{E}\!\left(\prod_{i=h}^1 B^{-1}_{i}A^{-1}_{i}B_{i}A_{i}\,\grpdot \univppt{z},p;X\!\right) \\
	&{=}\,\genfE{\univppt{z}}{X}\prod_{i=1}^\genus\monoA{i}{X}\monoB{i}{X}\monoA{i}{X}^{-1}\monoB{i}{X}^{-1}.
\end{aligned}
\end{equation}
On the other hand, the asymptotic behavior~\eqref{eqn:asymptotics} implies
\begin{equation}\label{eqn:CcycleAction2}
	\begin{aligned}
		\genfE{C\grpdot \univppt{z}}{X}
		&= \genfE{\univppt{z}}{X}\,e^{2\pi\iunit t}  \, .
	\end{aligned}
\end{equation}
Comparing eqs.~\eqref{eqn:CcycleAction} and \eqref{eqn:CcycleAction2}, one finds
\begin{equation}\label{eqn:monorel}
	\prod_{i=1}^\genus\monoA{i}{X}\monoB{i}{X}\monoA{i}{X}^{-1}\monoB{i}{X}^{-1} = e^{2\pi\iunit t} \, ,
\end{equation}
which is the echo of the presentation of $\fgp$ for the monodromy series.

To formulate an ansatz for a generating function of svhgPLs, we first want to cancel the factor $e^{-2\pi\iunit b_i}$, $i\,{\in}\,[\genus]$, when analytically continuing $\genfE{z}{X}$ around $\mathrm{B}_i$ (c.f.~\eqn{eqn:EmonoB}). As explained \hyperlink{target:T}{above}, this can be done by means of a trivialization $\bfunc(z,p;X)\,{\in}\,\funcs{\RSurfuniv}\otimes\cG\cU$, which is required to satisfy
\begin{equation}\label{eqns:gauge}
	\begin{aligned}
		\bfunc(p,p;X)&=1\, , \\
		\bfunc(A_j\grpdot z,p;X)&=\bfunc(z,p;X)\, , \\
		\bfunc(B_j\grpdot z,p;X)&=\bfunc(z,p;X)\,e^{2\pi\iunit b_j}.
	\end{aligned}
\end{equation}
For now, we just assume the existence of such a function, while describing a possible choice further \hyperlink{target:g}{below}.

Next, we want to define the automorphism $\iota\,{:}\,\cG\uea\,{\rightarrow}\,\cG\uea$ in order to formulate the ansatz~\eqref{eqn:ansatz-general} for $\genfE{z}{X}$. Explicitly, we set
\begin{equation}
	\begin{aligned}\label{eqn:svansatz}
		\svgenfE{\univppt{z}}{X} &\coloneqq \overline{\bfunc(\univppt{z},p;X_\SVaut)\,\genfE{\univppt{z}}{X_\SVaut}} \\
		&\qquad\times\left(\bfunc(\univppt{z},p;X)\,\genfE{\univppt{z}}{X}\right)^{-1} \, ,
	\end{aligned}
\end{equation}
where $X\,{\mapsto}\, X_\SVaut$ denotes an automorphism of $\free$ (resp.~$\uea$ by extension), which is to be determined such that the resulting function is rendered single-valued. Note that this ansatz encodes the map $\iota$ as the combination of the automorphism $X\,{\mapsto}\, X_\SVaut$ and complex conjugation\footnote{While complex conjugation is naively taken over from the constructions at genus zero and one, it should be possible to consider other involutions.}.

Exhibiting the monodromy properties of the generating function (cf.~\eqn{eqns:hgPTOs}), we can immediately formulate the \emph{higher-genus single-valued (hgsv) conditions} as 
\begin{equation}\label{eqns:svhg-cond}
	\begin{aligned}
		1&=\overline{\monoA{i}{X_\SVaut}}\,\monoA{i}{X}^{-1}=\overline{\monoB{i}{X_\SVaut}}\,\monoB{i}{X}^{-1}\, , \,\, i\in[\genus]\, , \\
		1&=e^{2\pi\iunit t}\,e^{2\pi\iunit\SVaut(t)} \, .
	\end{aligned}
\end{equation}
We note that the validity of the third condition is implied by the first two, which follows by direct computation making use of \eqn{eqn:monorel} and the requirement that $\monoA{i}{X_\SVaut}\,{\in}\,\cG\uea$, which is equivalent to demanding that $X\,{\mapsto}\, X_\SVaut$ is an automorphism of $\free$. 

Above, we derived a system of $2\genus$ independent equations for the $2\genus$ unknowns given by $X_\SVaut$, whose solution is equivalent to the statement of the function being single-valued, i.e.~$\svgenfE{z}{X}\,{\in}\,\funcs{\pRSurf}\,{\otimes}\,\cG\uea$. In order to argue that this system indeed has a solution, we use a general result by Brown~\cite{Brownhyperlogs} (see also \rcite[app.~B]{Baune:2025ndr} for a proof), implying that it suffices to realize that the matrix of coefficients
\begin{equation}
	\begin{pmatrix}
		\monoA{}{X}[a] & \monoA{}{X}[b] \\
		\monoB{}{X}[a] & \monoB{}{X}[b]
	\end{pmatrix} = \begin{pmatrix}
		\Id &  0 \\
		\tau & 2\pi\iunit\Id
	\end{pmatrix}
\end{equation}
is invertible. Here, e.g.~$\monoA{}{X}[a]$ refers to the matrix $(\monoA{j}{X}[a_i])_{i,j\in[\genus]}$, where $\monoA{j}{X}[a_i]$ denotes the coefficient of $a_i$ in $\monoA{j}{X}$. To compute the right-hand side, we used definitions~\eqref{eqns:hgPTOs} as well as the expansion~\eqref{eqn:exp_loworder}. Solving the hgsv conditions \eqref{eqns:svhg-cond} yields (to first order)
\begin{subequations}
	\begin{align}
		\SVaut(a_i) &= a_i + \ldots \, , \\
		\SVaut(b_i) &= -b_i - \sum_{j=1}^\genus\frac{Y_{ij}}{\pi}a_j + \ldots
	\end{align}
\end{subequations}
for all $i\,{\in}\,[\genus]$, where $Y_{ij} \texteq \Im(\tau_{ij})$.
Plugging this solution into the ansatz \eqref{eqn:svansatz} yields a generating function $\cL(z,p;X)$ of svhgPLs, whose coefficients obey shuffle relations due to $\cL(z,p;X)$ being group-like. An explicit choice of a trivialization $\bfunc(z,p;X)$ and the resulting svhgPLs are discussed next.

\smallskip
\noindent\textbf{\textit{Choosing a trivialization.}}\label{sec:svhgplDHS} In the above construction, we have not yet specified the function $\bfunc(z,p;X)$ (c.f.~\eqn{eqn:svansatz}). \hypertarget{target:g}To complete the description of svhgPLs, we describe a possible choice of trivialization $\bfunc(z,p;X)$ and investigate its relation to the hgPLs constructed in \rcite{DHoker:2023vax}.

In \rcite{DHoker:2025szl}, a function $\gauge{\univpt{z}}{X}\,{\in}\,\funcs{\RSurfuniv}\,{\otimes}\,\cG\uea$ and an automorphism\footnote{The automorphism $X\,{\mapsto}\,X_{\DHSaut^{-1}}$ generally depends on the puncture $p$ (and on the choice of preimage $\univpt{p}$)~\cite{DHoker:2025szl}. However, to avoid cluttering, we will omit this dependence in the following.} $X\,{\mapsto}\,X_{\DHSaut^{-1}}$ of $\free$ (resp.~$\uea$ by extension) have been constructed such that $\gauge{z}{X_\DHSaut}^{-1}$ serves as a valid choice\footnote{Another possibility of trivialization has been constructed in \rcite{EZ1}, which in contrast features poles of higher orders while maintaining holomorphicity.} for the trivialization $\bfunc(z,p;X)$, i.e.~$\gauge{z}{X_\DHSaut}^{-1}$ satisfies the properties~\eqref{eqns:gauge}. Furthermore, notice that it has been shown in \rcite{DHoker:2025szl} that
\begin{align}
	\cdhs{\univpt{z}}{X} &= \gauge{\univpt{z}}{X}^{-1}\,\cenr{\univpt{z}}{X_{\DHSaut^{-1}}}\,\gauge{\univpt{z}}{X} \notag\\
	&\quad- \gauge{\univpt{z}}{X}^{-1}\,\dd\gauge{\univpt{z}}{X} \, ,
\end{align}
where $\cdhs{z}{X}\,{\in}\,\diffs{p}{\RSurf}\,{\otimes}\,\free$ is a differential form arising from a flat connection constructed by D'Hoker--Hidding--Schlotterer in \rcite{DHoker:2023vax}. It admits a simple pole with residue $t$ at $z\texteq p$ and, in contrast to $\cenr{\univpt{z}}{X}$, is real-analytic and single-valued. Defining
\begin{equation}\label{eqn:dhsPLs}
	\genfDHS{\univppt{z}}{X} \coloneqq \gauge{\univppt{z}}{X_\DHSaut}^{-1}\,\genfE{\univppt{z}}{X},
\end{equation}
we correspondingly obtain the differential equation
\begin{equation}
	\hspace{-1ex}\label{eqn:DHS-KZB}\dd \genfDHS{\univppt{z}}{X} = \cdhs{\univppt{z}}{X_\DHSaut}\,\genfDHS{\univppt{z}}{X} ,
\end{equation}
as well as the asymptotics $\genfDHS{\univppt{z}}{X}\,e^{-t\log(\univppt{z})}\,{\overset{z\rightarrow p}{\sim}}\,1$. The shuffle relations are again encoded in the fact that $\genfDHS{z}{X}$ is group-like. Accordingly, given this choice of trivialization, our generating function of svhgPLs can be rewritten as 
\begin{equation}\label{eqn:svDHS}
	\svgenfE{z}{X} = \overline{\genfDHS{z}{X_\SVaut}}\,\genfDHS{z}{X}^{-1} \, ,
\end{equation}
which consequently links to the hgPLs constructed in \rcite{DHoker:2023vax}. 

\smallskip
\noindent\textit{\textbf{Reduction to genus one.}} From \rcite[rmk.~4.10]{DHoker:2025szl}, we conclude that $g(z,0;X_\xi)^{-1}\,{=} \exp(2\pi\iunit\frac{\Im(z)}{\Im(\tau)}b)$ upon specializing to genus one, where the automorphism $\xi$ is now explicitly given by $a\,{\mapsto}\, a\,{+}\,\frac{\pi}{\Im(\tau)}b$ and $b\,{\mapsto}\, b$. Moreover, it can be verified that $\mathcal{K}_\mathsf{DHS}(z,0;X_\xi)$ exactly coincides with the connection introduced by Brown--Levin in \rcite{BrownLevin} at genus one. Therefore, the function $\mathrm{L}_\mathsf{DHS}(z,0;X)$ defined in \eqn{eqn:dhsPLs} reduces (up to a conventional sign and normalization) to the generating function $\Gamma(z|\tau)$ from \rcite[eqs.~(3.4),~(3.5)]{Baune:2025ndr} when reducing to genus one. Therefore, we can write the generating function of svhgPLs~\eqref{eqn:svDHS} in terms of $\Gamma(z|\tau)$ at genus one, thereby establishing a link to the constructions from \rcites{Baune:2025ndr,Schlotterer:2025qjv}. Indeed, when explicitly evaluating \eqn{eqn:svDHS} to second order at genus one, we obtain
\begin{equation}\label{eqn:genusone}
	\mathcal{L}(z,0;X) \overset{\genus=1}{=} 1+ \!\left(-2\Re(\ell_{ba})+2\pi\frac{\Im(z)^2}{\Im(\tau)}\right)\![b,a] + \ldots,
\end{equation}
which coincides with the result from \rcite[eq.~(C.4)]{Baune:2025ndr}.

\section{Examples and Open Questions}

\noindent\textbf{\textit{Comparison to the Arakelov Green's function.}} The Arakelov Green's function $\ara{z}{p}$ (see e.g.~\rcites{Arakelov_1974,FaltingsArakelov} for a mathematical account and e.g.~\rcite{DHoker:2017pvk} for its applications in string theory) is uniquely characterized as the smooth, symmetric real-valued function on $\RSurf\,{\times}\, \RSurf\,{\setminus}\,\Delta$, where $\Delta$ is the diagonal, which adheres to the properties (using the conventions\footnote{With $\kappa(z)\texteq\frac{\iunit}{2\genus}\sum_{j,k=1}^\genus (Y^{-1})_{jk}\omega_j(z)\wedge\overline{\omega}_k(z)$ being the normalized volume form on $\RSurf$.} of \rcite[eq.~(2.24)]{DHoker:2025szl})
\begin{equation}\label{eqn:Gdef}
	\pd_z\pd_{\bar z}\ara{z}{p}=2\pi\iunit\left(\delta(z,p)-\kappa(z)\right),\ \ \int_\RSurf\kappa(z)\ara{z}{p}=0.
\end{equation}
At genus one, we find the Arakelov Green's function as coefficient of the word $ba$ (or $ab$ up to a minus sign) of $\cL(z,0;X)$ in \eqn{eqn:genusone}. At higher genera, we thus similarly expect $\ara{z}{p}$ to be related to the coefficients $\cL(z,p;X)[a_ib_i]$, $i\,{\in}\,[\genus]$. Since $\ara{z}{p}$ is real-valued, we consider the real parts of $\cL(z,p;X)[a_ib_i]$, which are independently single-valued. When additionally summing over $i\,{\in}\,[\genus]$, we obtain
\begin{align}
	-h\cF(z,p)\coloneqq&\Re(\cL(z,p;X)[a_ib_i])\notag\\
	=&\,2\Re(\ell_{b_ia_i}\!(z,p))\notag\\
	&- 2(Y^{-1})_{lm}\Im(\ell_{a_m}\!(z,p))\Re\!\left(\ell_{b_ia_i}\!(B_l\grpdot p,p)\right)\notag\\
	&-2\pi(Y^{-1})_{im}\Im(\ell_{a_m}\!(z,p))\Im(\ell_{a_i}\!(z,p))\notag\\
	&+2\pi(Y^{-1})_{im}\Im(\ell_{a_m}\!(z,p))Y_{ii}\Big],
\end{align}
with implicit summations over repeated indices $i,l,m$ and the coefficients written according to the notation introduced in \eqn{eqn:sol}. Both $\ara{z}{p}$ and $\cF(z,p)$ locally expand as ${-}\log|z{-}p|^2$ plus regular terms (c.f.~\rcite{DHoker:2025szl}), when $z$ is close to $p$. Therefore, the difference $\cH(z,p)\texteq\ara{z}{p}\,{-}\,\cF(z,p)$ constitutes a well-defined smooth function on all of $\RSurf\,{\times}\,\RSurf$. Moreover, one can check that $\cF(z,p)$ satisfies the same differential equation as $\ara{z}{p}$ (c.f.~\eqn{eqn:Gdef}) in both arguments $z$ and $p$. Hence, $\cH(z,p)$ is a global harmonic function in $z$ as well as $p$, and application of the maximum principle~\cite{FarkasKra1992} implies that $\cH(z,p)\texteq\cH$ is a (real) constant. Thus we conclude that $\cF(z,p)\texteq\ara{z}{p}\,{+}\,\cH$. The constant $\cH$ can be determined by the normalization (c.f.~\eqn{eqn:Gdef}) since
\begin{equation}
	\cH = \int_\RSurf\kappa(z)\cH = -\int_\RSurf\kappa(z)\cF(z,p) \, .
\end{equation}

\noindent\textbf{\textit{Connection to known objects and constructions.}}  At genus one, the Arakelov Green's function as well as the elliptic Bloch--Wigner dilogarithm are the most widely known single-valued objects. While the link of our construction to the Arakelov Green's function was discussed above, a general notion of defining a higher-genus analog of the genus-$h$ generalization of the Bloch--Wigner dilogarithm remains elusive at this point. 

Furthermore, it is unclear in what sense the monodromy series defined in \eqn{eqns:hgPTOs} can be regarded as a realization of higher-genus associators in the sense of \rcites{gonzalez2020surfacedrinfeldtorsorsi,taniguchi2026drinfeldassociatorskashiwaravergneassociators}.

It would also be interesting to see, how our construction is related to the considerations in \rcite{Pokraka:2025zlh}.

Whenever a flavor of polylogarithmic functions is defined, the question about their special values is immediate. In the context of svhgPLs, one should be able to find a suitable definition of single-valued higher-genus mutliple zeta values and relate them to modular graph functions.  

Finally, the physics intention fueling the current investigation is twofold: while finding a canonical language for expressing results of single-valued scattering amplitudes is a virtue in its own right, the formalism explored in this article should constitute a good starting point to investigating and constructing KLT-like or double-copy representations for --- in particular --- higher-genus closed-string scattering amplitudes. 

\vspace*{0.5cm}
\begin{acknowledgments}
\noindent\textbf{Acknowledgments.}
We are grateful to Egor Im and Federico Zerbini for various discussions and work on related projects. 
KB and JB would like to thank the ESI Vienna for hospitality during the thematic program ``Amplitudes and Algebraic Geometry 2026''.
The work of all authors is partially supported by the Swiss National Science Foundation through the NCCR SwissMAP.
\end{acknowledgments}


\bibliography{svhgmpls}
\bibliographystyle{apsrev4-1}

\end{document}